# PAM-4 Transmission at 1550nm using Photonic Reservoir Computing Post-processing

Apostolos Argyris[1], Julián Bueno[1,2], and Ingo Fischer[1]

[1]Instituto de Física Interdisciplinar y Sistemas Complejos IFISC (CSIC-UIB), Campus UIB, 07122, Palma de Mallorca, Spain
[2]now with Institute of Photonics, SUPA Department of Physics, University of Strathclyde, TIC Centre, 99 George Street, Glasgow G1 1RD, U.K. Kingdom

Corresponding authors: Apostolos Argyris (e-mail: apostolos@ifisc.uib-csic.es), Ingo Fischer (e-mail: ingo@ifisc.uib-csic.es)

This work was supported by the Ministerio de Economía y Competitividad and FEDER via project IDEA (TEC2016-80063-C3) and by the European Union's Horizon 2020 research and innovation programme under the Marie Skłodowska-Curie project CENTURION (contract 707068).

**ABSTRACT** The efficacy of data decoding in contemporary ultrafast fiber transmission systems is greatly determined by the capabilities of the signal processing tools that are used. The received signal must not exceed a certain level of complexity, beyond which the applied signal processing solutions become insufficient or slow. Moreover, the required signal-to-noise ratio of the received signal can be challenging, especially when adopting modulation formats with multi-level encoding. Lately, photonic reservoir computing (RC) – a hardware machine learning technique with recurrent connectivity – has been proposed as a post-processing tool that deals with deterministic distortions from fiber transmission. Here we show that RC post-processing is remarkably efficient for multilevel encoding and for the use of very high launched optical peak power for fiber transmission up to 14dBm. Higher power levels provide the desired high signal-to-noise ratio (SNR) values at the receiver end, at the expense of a complex nonlinear transformation of the transmission signal. Our demonstration evaluates a direct fiber communication link with 4-level pulse amplitude modulation (PAM-4) encoding and direct detection, without including optical amplification, dispersion compensation, pulse shaping or other digital signal processing (DSP) techniques. By applying RC post-processing on the distorted signal, we numerically estimate fiber transmission distances of 27km at 56Gb/s and of 5.5 km at 112Gb/s data encoding rates, while fulfilling the hard-decision forward error correction (HD-FEC) bit-error-rate (BER) limit for data recovery. In an experimental equivalent demonstration of our photonic reservoir, the achieved distances are 21km and 4.6km respectively.

**INDEX TERMS** Machine learning, nonlinear dynamics, optical signal processing, reservoir computing, semiconductor lasers.

## I. INTRODUCTION

Building cost-efficient and low-complexity systems using intensity modulation / direct detection (IM/DD) schemes at 1550nm for data center, access, and metro communications is subject to severe limitations [1,2]. Data recovery at the communicating end must be capable of dealing with chromatic dispersion, the square-law photodetection nonlinearity, as well as with Kerr-induced nonlinearity in the presence of high-power optical signals [3]. In all implementations of this type of communication, the use of digital signal processing (DSP) – offline or real-time – has been essential to mitigate transmission and detection impairments and to perform equalization tasks [4-8]. Even in the absence of physical dispersion compensation, standard single mode fiber (SSMF) transmission lengths up to 300km, using PAM-4 encoding at 1550nm, have been reported with the use of optical amplification [9-11]. Configurations without introducing optical amplification also perform remarkably well in a short transmission range, by using DSP techniques such as pre-compensation dispersion at the transmitter, decision feedback equalizers and maximum likelihood sequence estimations. For example, a transmission distance of 26.4km has been reported at 56Gb/s in a single 1550nm SSMF channel [12]. Pushing such systems to operate at the limit by optimizing the bandwidth-distance product increases the demands for technologically advanced signal processing. The aforementioned works optimize DSP, by considering the signal properties after SSMF fiber transmission. Yet, this is not the only strategy to optimize the performance of the fiber transmission system. Machine learning (ML) and neural network (NN) techniques start to show their potential in addressing problems related to fiber-based transmission systems [13-16]. Lately, transmission impairments have also been mitigated using various ML and NN approaches [17-20]. The latter may achieve similar performance to DSP solutions which have been specifically designed to address a given transmission topology. Hardware-friendly ML techniques, such as RC [21], represent attractive alternatives to the conventional ML approaches. Implementations based on photonic RC have set a new framework for solving diverse classification and equalization tasks [22-30]. Very recently, RC based on a semiconductor laser with optical feedback has been proposed by the authors to address signal recovery in optical communication systems [31,32], while later other RC topologies have also been tested for signal equalization from optical transmission systems [33,34].

While in [31] the evaluated transmission systems had not so strict SNR requirements for signal detection, due to a 2-level encoding at lower bit rates, here we demonstrate the potential of photonic RC in a parameter space well beyond the one used in classical implementations of optical communication systems. We focus on a specific communication system that currently applies to short-reach passive networks. This is the 4-level amplitude modulation encoding (PAM-4) at bit rates of 56Gb/s and 112Gb/s. The deployment of this system has specific requirements for high optical SNR (OSNR) signal detection. Thus, in order to obtain high OSNR and transmit at the longest distance possible without any amplification, we propose to increase the launched optical power for transmission at levels that are not conventional. Such high power induces strong nonlinear signal distortion that is not easy to cope with using conventional digital signal processing. In this work we demonstrate that the highly nonlinear transformation that the signal undergoes after transmission at launched optical peak power levels up to 14dBm is remarkably equalized by the photonic RC.

In the next section we describe the transmission system configuration that provides the nonlinearly distorted signals and the subsequent RC post-processing technique. In section III, we evaluate the data recovery performance versus the reservoir operating point, the properties of the transmission signal in terms of OSNR and launched optical peak power and the RC training conditions. In section IV, we validate the numerical findings in an experiment, feeding the detected transmission signals into a fiber-based, photonic reservoir. The operating conditions of the implemented reservoir are determined by the optimization mapping presented in [32]. Finally, section V summarizes the findings of this study.

## II. SYSTEM CONFIGURATION

The proposed methodology is illustrated in a flow chart in Fig. 1. A communication channel is simulated, while the decoding process is assisted by a photonic reservoir that performs a physical nonlinear transformation with memory properties. The output reservoir signal – along with the initial encoded data-stream information – is used to train a linear regression algorithm. This allows us to evaluate independent data sets of the communication channel. Finally, a comparator between the encoded and the decoded data sets evaluates the error rate level of the communication channel.

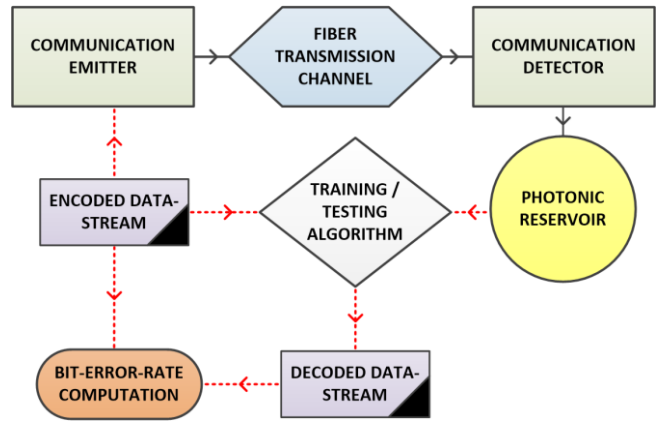

**FIGURE 1.** Flow chart of the proposed topology for post-processing the received signals from the communication channel using a photonic reservoir. Black (straight) lines show physical system connectivity. Red (dashed) lines show computational connectivity.

### A. FIBER TRANSMISSION

We numerically simulate a simple physical transmission SSMF channel (Fig. 2a) of length $L$ via the coupled nonlinear Schrödinger equations (CNLSE). We consider a PAM-4 IM/DD transmission system designed for 1550nm at data rates of $R_1$=56Gb/s and $R_2$=112Gb/s and we examine the simplest possible structure for point-to-point transmission. Thus, we do not consider any optical amplification, dispersion compensation, filtering for pulse shaping or traditional DSP techniques for equalization and nonlinear mitigation that require a prior knowledge of the transmission channel. The model considers two orthogonal polarization modes and stimulated Brillouin and Raman scattering, while inter-channel nonlinear effects (such as cross-phase modulation and four-wave mixing) do not apply for our single-channel consideration [35]. A distributed feedback (DFB) semiconductor laser (SL) emits at 1550nm with a relative intensity noise (RIN) set to -150dB/Hz, while an equidistant 4-level amplitude encoding of the data stream is applied through a Mach-Zehnder modulator (MZM) operating in the linear regime. Bandwidth-limited photodetection is assumed through a PIN receiver with transimpedance gain (TIA) that includes thermal and shot noise effects, with a frequency cutoff at 0.7 of the data encoding bit rate. Thus, optical noise in our system originates entirely from the laser source. The rest of parameters used for the transmission simulation are shown in table 1. A critical parameter in our study is the high launched peak optical power for transmission, which also results in a received signal with high OSNR. For the different cases of the system data rates $\{R_1, R_2\}$ we consider such transmission lengths $\{L_1, L_2\} = \{27km, 5.5km\}$ that lead to a performance below the BER HD-FEC limit ($3.8 \cdot 10^{-3}$) [36] after the proposed RC processing. In absence of any post-processing, signal recovery exhibits a BER value as high as 0.2 for both bit rates.

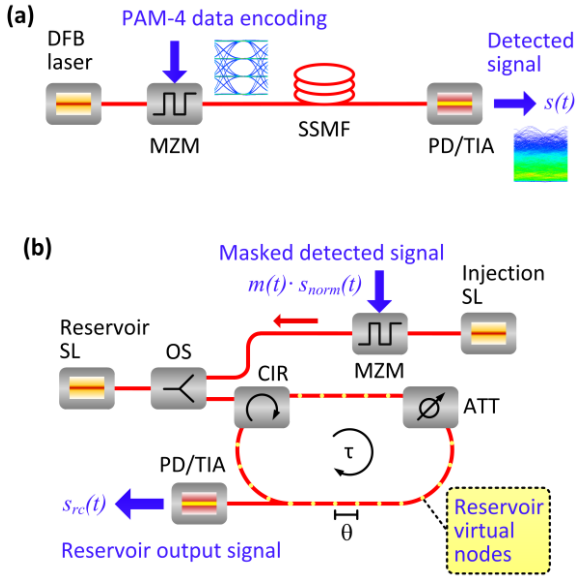

**FIGURE 2.** (a) Configuration of a PAM-4 IM/DD transmission system without dispersion compensation, optical amplification or DSP. The eye-diagrams refer to the detected signal after back-to-back operation and after SSMF transmission of 27km, at 56Gb/s. (b) Photonic reservoir concept for signal post-processing based on a SL and a single feedback delay line. OS: optical splitter, CIR: optical circulator.

### B. RESERVOIR COMPUTING POST-PROCESSING

In the adopted reservoir topology, the optical signal from transmission is photodetected – $s(t)$ – before being introduced to the photonic reservoir. This optoelectronic conversion stage abolishes the polarization and phase properties of the optical signal before entering the reservoir, allowing only the intensity information to be processed. The photonic reservoir, as an independent optical system, operates in its own polarization state and, therefore, is robust to any polarization changes in the transmission channel. An attractive consideration for future investigations is an all-optical reservoir input stage, where the optical signal from transmission is fed directly into the photonic reservoir. However, such an approach imposes additional challenges for the implementation, such as continuously preserving the same polarization state between the received optical signal from transmission and the operating photonic reservoir. A coherent implementation of the reservoir, extending the concept for coherent encoding and detection schemes is beyond the scope of this manuscript.

Before processing the detected signal $s(t)$ from transmission, it is normalized – $s_{norm}(t) \in [0,1]$ – and multiplied by a random mask sequence of random values $m(t) \in [0,1]$. The masking is applied to every single baud pattern. Its role in this approach is to increase the dimensionality of the state-space representation of the signal to be processed. The masked signal $m(t) \cdot s_{norm}(t)$ is then introduced into the photonic reservoir as an optical signal via a MZM operating near the linear regime. The photonic reservoir represents a time-multiplexed photonic network, as originally introduced in [22]. In the proposed implementation, the input signal $m(t) \cdot s_{norm}(t)$ is nonlinearly transformed and collected at the output of the reservoir ($s_{rc}(t)$). From this output one can extract ultra-fast transient states in order to train a linear classifier and obtain the reconstructed initial transmitted data stream.

The considered implementation of the photonic reservoir is shown in Fig. 2b. It is formed by a reservoir SL and an optical delay line of $\tau$ that introduces recurrent connectivity between the virtual nodes defined in the optical delay path. We define $N=32$ equidistant virtual nodes and thus $N$ transient states with spacing $\theta = \tau / N$ along the optical delay. The number of virtual nodes dictates also the scale of oversampling we apply on $s(t)$, as well as the dimension of $m(t)$, so that every transient state emerges from a masked input value. Every baud pattern in the transmission simulation is described by 8 samples. Thus, an oversampling of 4 is applied to this signal, while the mask vector consists of 32 random values. Consequently, the dimensionality of the processed signal is increased due to the reservoir transformation. The number of samples as well as the number of virtual nodes used in this demonstration is less than the ones presented in the experimental RC topology of [31], where the dimensionality of the input was preserved during the reservoir transformation. From the two distinct time scales defined in the processing methodology – the duration of one pattern (baud) and the duration of time delay of the reservoir – one can easily deduct the induced speed penalty. Since each baud pattern (2 bits of information in one time-frame unit) is assigned to the $N$ virtual nodes of the reservoir's time delay, this time-multiplexing reservoir implementation is an offline process. The speed penalty (SP) of the processing step is: $SP = (R/2) \cdot \tau$, where $R$ is the encoding bit rate and $\tau$ is the time delay of the reservoir's optical feedback loop. For example, in the case of $R_1$ and $\tau = 0.8ns$, $SP$ is equal to 22.4.

The optical feedback received by the reservoir is controlled by tuning the internal attenuation (ATT). Thus, the nonlinear transformation originates from both reservoir SL and time-delayed feedback. The numerical model follows the Lang-Kobayashi rate equations of a SL with time-delayed feedback, with an additional optical injection dynamical term with frequency detuning $\Delta f = f_{inj} - f_r$. An analytical description of the model with optical feedback and optical injection of frequency detuned signals can be found in [37]. The slowly varying electrical field amplitude

$E_r(t)$ corresponding to the optical emission of the response SL is calculated using the following equations:

$$\frac{dE_r(t)}{dt} = \frac{1}{2}(1+ja)[G_r(t) - t_{ph}^{-1}] \cdot E_r(t) + \frac{k_f}{t_{in}} \cdot E_r(t-\tau)e^{j\omega_0\tau} + \frac{k_{inj}}{t_{in}} \cdot E_{inj}(t)e^{-j\Delta\omega t} + \sqrt{D} \cdot \xi(t) \quad (1)$$

$$\frac{dN_r(t)}{dt} = \frac{I}{q} - \frac{N_r(t)}{t_s} - G_r(t) \cdot |E_r(t)|^2 \quad (2)$$

$$G_r(t) = g_n \cdot [1 + s|E_r(t)|^2]^{-1} \cdot [N_r(t) - N_0] \quad (3)$$

$$E_{inj}(t) = E_{inj,0} \cdot \left[\frac{1}{2} + m(t) \cdot a(t)\right] \quad (4)$$

The signal to be processed by the reservoir is inserted by modulating an independent optical carrier (injection SL) through the linear operation of a MZM and its optical output is injected into the reservoir SL. The injected electrical field to the reservoir is of the form of (4), with an average amplitude of $E_{inj,0}$. A Gaussian white noise term $\xi(t)$ is included for the electrical field with amplitude $D = 3\text{ns}^{-1}$. $q$ is the electron charge. The rest of parameters used for simulating the reservoir are summarized in table I.

TABLE I
PARAMETERS FOR THE NUMERICAL ANALYSIS

| Parameters for the transmission link | |
|---|---|
| Transmission loss coefficient | 0.2 dB/km |
| Chromatic dispersion coefficient | 17 ps/(nm·km) |
| Nonlinear index of refraction | $2.6 \cdot 10^{-20}$ m²/W |
| Photoreceiver responsivity | 0.9 A/W |
| Transimpedance amplifier gain | 4 to 16 dB |
| Differential group delay | 0.2 ps/km |
| Effective area of SSMF core | 80 μm² |
| Launched optical peak power | -8 to +14 dBm |
| **Parameters for the photonic reservoir** | |
| Frequency detuning | $\Delta f$: -50 to +50 GHz |
| Angular frequency detuning | $\Delta\omega = 2\pi \cdot \Delta f$ |
| Reservoir SL bias current | $I = 15.3 \cdot 10^{-3}$ A |
| Reservoir SL threshold current | $I_{th} = 15.37 \cdot 10^{-3}$ A |
| Linewidth enhancement factor | $a = 3$ |
| Gain coefficient | $g_n = 1.2 \cdot 10^{-5}$ ns⁻¹ |
| Gain saturation coefficient | $s = 5 \cdot 10^{-7}$ |
| Carrier number at transparency | $N_0 = 1.5 \cdot 10^8$ |
| Carrier lifetime | $t_s = 2$ ns |
| Reservoir SL roundtrip time | $t_{in} = 10^{-2}$ ns |
| Photon lifetime | $t_{ph} = 2 \cdot 10^{-3}$ ns |
| Optical angular frequency | $\omega_0 = 2\pi c/\lambda_0$ |
| Injection parameter | $k_{inj} = 0.15$ |
| Feedback parameter for the reservoir SL | $k_f$: 0 to 0.2 |
| Time delay of optical feedback loop | $\tau$: 0.8ns or 1.6ns |
| Average amplitude of injected electrical field | $E_{inj,0} = 100$ |

At the final readout stage, a weighted summation of all virtual nodes' responses $s_{rc}(t)$ is performed to predict each baud value. We determine the weights (predictor variables) via an offline linear (ridge) regression (LR) algorithm. We consider streams of $2^{17}$ baud patterns ($2^{18}$ bits) in all studied cases, 75% of which are used for training and 25% for cross-validation to monitor the training efficiency. Independent data streams of equal length are then used as test sets in order to evaluate the system's performance. The time-varying signals, along with their spectral profiles, are shown in Fig. 3, at different stages when passing through the system: the launched optical signal in the fiber transmission link (upper, black line), the detected signal after fiber transmission (middle, red line) and the randomly masked signal from detection that is fed into the reservoir, after offline time stretching (lower, blue line).

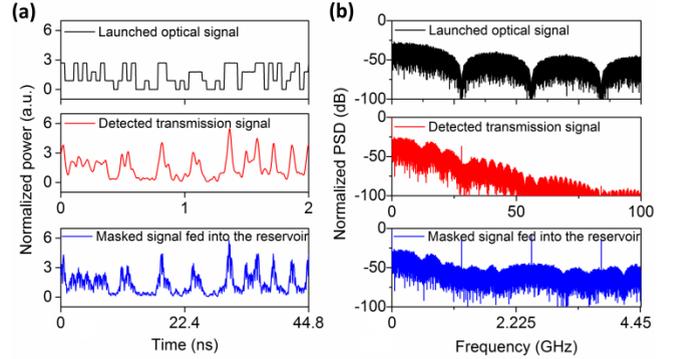

**FIGURE 3.** (a) Segment of a PAM-4 time-series at $R_1$ and (b) corresponding power spectra of the: 10dBm launched optical signal in the fiber transmission link (upper, black), detected signal after $L_1$ SSMF transmission (middle, red) and randomly masked detected signal that is fed into the reservoir, after offline time stretching (lower, blue). Time-series are normalized with standard deviation σ=1 for visualization.

### III. RESULTS AND DISCUSSION

Semiconductor lasers with optical feedback, which are in addition subject to an external continuous optical injection, are systems of great interest due to the complex dynamics they exhibit [37]. When the injection is dynamical, the response of the system becomes even more complex and potentially high-dimensional and significantly affected by the properties of the injected optical signal. The reservoir system that we are exploring here exhibits such attributes. Therefore, for a given parameter set, the reservoir will exhibit various dynamical responses. A systematic dynamical analysis of such systems has not yet been provided in literature. Nevertheless, there are other – statistical – features of the responses that can provide some useful information regarding the system operation. One of them is the signal-to-noise ratio (SNR) of the reservoir output. By combining the attributes of the SNR and the final BER performance of the reservoir classifier, one can categorize the reservoir's behavior into three different classes, depending on the frequency detuning of the injected light into the reservoir and its feedback strength: (a) a fully injection-locked operation, where the reservoir applies a rather small signal transformation, (b) a partially injection-locked operation where the reservoir response is a stronger nonlinear transformation of the injected signal, and (c) and an unlocked operation where the reservoir response is less consistent to the injected signal [38]. In the presence of strong optical feedback, the chaotic operation of the reservoir becomes dominant. The latter operating regime shows inconsistent responses to a given input and is not appropriate for computation tasks.

In Fig. 4a, we show the SNR of the reservoir output signal versus the frequency detuning $\Delta f$ and the feedback ratio $k_f$, for $\{R_1, L_1\}$ and $\theta$ = 50ps. In the full injection locked regime a very high SNR of the reservoir's response is recorded. Especially around the region of $\Delta f$ = -10GHz, the SNR exceeds 40dB. This is observed for low and moderate feedback values. For high feedback conditions, chaotic emission dominates and SNR becomes independent of the frequency detuning of the injected signal. In Fig. 4b we contrast the classification efficiency of the RC in terms of BER in an equivalent map. For full injection locking operation ($\Delta f$ = -10GHz), even though SNR is high, RC underperforms. Reservoir emission with high SNR can also be obtained for strong feedback conditions ($k_f$ > 0.15), where chaotic dynamics is dominating. Also under these conditions, the decoding performance of the reservoir is poor. Feedback-induced chaotic dynamics prevails over the properties of the injected signal that is to be processed. In contrast, completely unlocked operation of the reservoir is obtained for large $|\Delta f|$ values. Such conditions result in much smaller dynamical transients being induced in the reservoir. Consequently, the low SNR reservoir responses deteriorate the classification performance. For partial dynamical injection locking, however, we observe that RC significantly outperforms our benchmark, which is obtained by applying the linear regression directly to the transmission output signal (Fig. 4b, $BER_{LR}$ = 0.037, blue dashed line). For $\Delta f$ = 0GHz and $k_f$ = 0.05 the trained RC provides a signal recovery with $BER_{RC,min}$ = 2·$10^{-3}$, lower than the required hard-decision forward error correction threshold for error free decoding. In fact, there is a finite parameter region in the presented map where the obtained BER values are below the HD-FEC threshold (Fig. 4b, $BER_{HD-FEC}$, white dashed line). For the partial locking conditions, the SNR of the reservoir output signal is sufficient for the classification task, while the reservoir memory is maximized, as found in [38].

The temporal characteristics of the transient responses used for the reservoir computation affect the training performance. After endorsing a sufficient number of virtual nodes per baud pattern – in our case $N$ = 32 – shorter reservoir delays can also be considered by reducing the virtual node separation. We reduce $\theta$ to 25ps and repeat the previous investigation (Fig. 4c,d). The mapping of the SNR (Fig. 4c) and the RC BER (Fig. 4d) performance is comparable. However, the HD-FEC limit is now achieved in a narrower regime of operating conditions, with a $BER_{RC,min}$ = 2.5·$10^{-3}$, slightly higher than for $\theta$ = 50ps. This virtual node spacing is rather at the edge of the bandwidth limitations for such systems, establishing also a minimum speed penalty for the time-multiplexed reservoir processing.

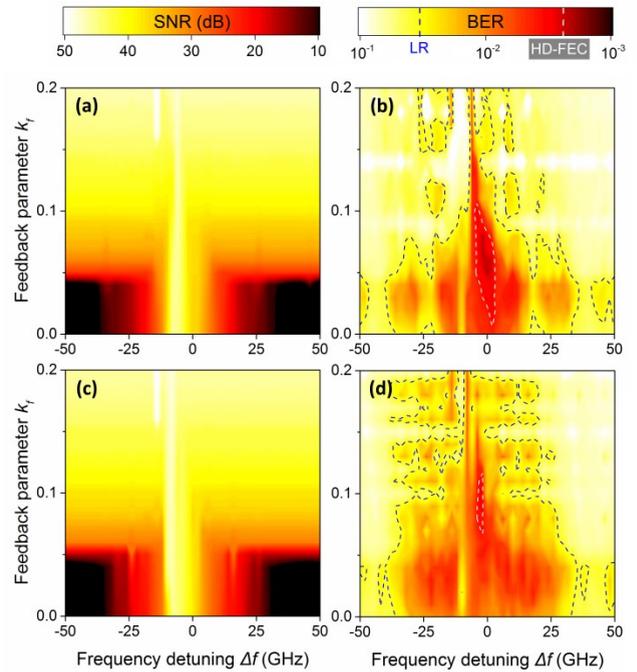

**FIGURE 4.** (a) SNR of the reservoir response signal and (b) BER evaluation of a PAM-4 56Gb/s test time-series after reservoir computing post-processing, versus the frequency detuning $\Delta f$ between the reservoir laser and the injection laser and the optical feedback parameter $k_f$. The reservoir delay is $\tau$=1.6ns ($\theta$=50ps). $BER_{LR}$=0.037 and $BER_{HD-FEC}$=3.8·$10^{-3}$. (c),(d): Same as (a),(b) but for $\tau$=0.8ns ($\theta$=25ps).

As discussed in this section, a systematic stability and dynamical analysis of this nonlinear system with dynamical external injection is lacking and hard to obtain. Moreover, an easy mapping of the reservoir's nonlinear transformation is not possible, since it is not static and high-dimensional. However, we can visualize the signal transformation in time-domain, for various dynamical operating regimes of the reservoir. For a given reservoir input $m(t)·s_{norm}(t)$ (Fig. 5a), we record the corresponding output of the reservoir $s_{rc}(t)$ for different operating conditions and corresponding dynamical regimes. When operating the reservoir in a full injection-locking regime (i.e. for $\Delta f$ = -10GHz and $k_f$ = 0.05) the signal is only weakly transformed (Fig. 5b). The output is highly correlated with the input (Pearson correlation of 0.92) and thus the reservoir decoding performance is limited. When operating the reservoir with high optical feedback (i.e. for $\Delta f$ = 0GHz and $k_f$ = 0.2), chaotic emission is induced. The reservoir is dominated by its internal dynamics and less by the input information (Fig. 5c), resulting in an inconsistent transformation. In this case, the Pearson correlation between the input and the output of the reservoir is dropping to 0.08. Consequently, no efficient decoding processing can be obtained. According to Fig. 4b, the operating regime where the optimal decoding is found lies on the boundaries of the injection-locking regime, in presence of moderate optical feedback. For $\Delta f$ = 0GHz and $k_f$ = 0.05, the reservoir output becomes less correlated with the applied input (Fig. 5d), compared to the full-locking condition, with a Pearson correlation of 0.87. In this operating regime, the photonic reservoir response remains sufficiently consistent, while still providing a pronounced nonlinear transformation.

By evaluating the results of Fig. 4 and Fig. 5, we conclude that a substantial BER performance improvement has only been obtained in the dynamical regime of partial injection locking. The reasoning is that only in this regime we fulfill fundamental attributes of reservoir computing, such as transformation consistency and internal fading memory. Regarding transformation consistency, similar inputs should be transformed – even nonlinearly – to similar outputs. This condition is not fulfilled when the optical feedback is too strong (chaotic operation). Fading memory is also important to the specific processing task that we study. Extended memory is also aligned with the dynamical regime of partial injection locking [38]. In a nonlinear transmission channel, previous bit responses affect the shape of the bit pattern on the current timeframe. These responses are preserved in the reservoir for several roundtrips $\tau$ in the optical cavity before they fade away, affecting the nonlinear transformation of the current bit pattern. Some bit error rate improvements have also been found in small islands of the parameter space $\{\Delta f, k_f\}$ that appear to be beyond the partial injection locking condition. Even though we cannot provide a strict boundary definition of the dynamical regimes for this system, we assume that in these islands the counterbalance effects among the input nonlinear transformation, its transformation consistency and the fading memory result in a slight improvement compared to a trained system with just the input.

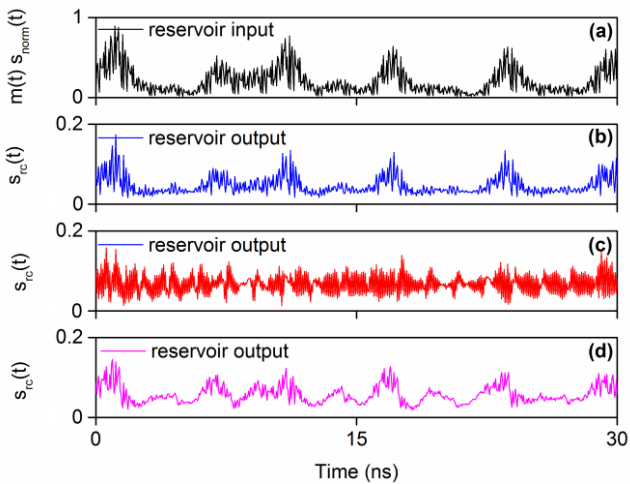

**FIGURE 5.** (a) Sample of timeseries of the normalized masked input that enters the photonic reservoir and the corresponding output of the reservoir, for different operating conditions and dynamical regimes: (b) $\Delta f$=-10GHz and $k_f$=0.05, for full injection-locking, (c) $\Delta f$=0GHz and $k_f$=0.2, for chaotic emission, and (d) $\Delta f$=0GHz and $k_f$=0.05, for partial injection-locking. $\tau$=1.6ns ($\theta$=50ps).

In the classification performance shown in Fig. 4b and Fig. 4d, we considered 20 taps of the reservoir response, since chromatic dispersion extends the pattern correlations to neighboring timeframes. 20 taps in our notation means that besides the reservoir response on a given input baud pattern, the responses of the 10 previous and the 10 next baud patterns are considered for the training process as well. Consequently, $21 \cdot N = 672$ transient states are used to optimize equal weights and train the classifier. The dependence of the BER performance on the number of taps is shown in Fig. 6. We find that the consideration of more than 20 taps leads to no further improvement of the BER performance, introducing only uncorrelated information from distant baud patterns. For each tapping condition, the optimal operating conditions for the reservoir were selected such that the lowest BER value for the evaluated test sequences was obtained. The latter approach is followed for all subsequent BER estimations.

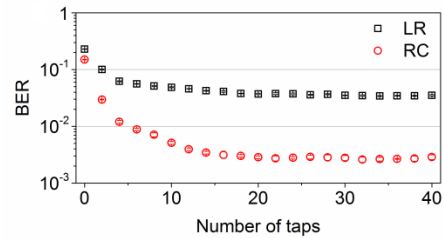

**FIGURE 6.** BER evaluation of a PAM-4 56Gb/s test time-series, after 27km of fiber transmission, versus the number of taps used for training the classifier. The training is performed directly on the detected signal from transmission output (black rectangles) and on the output of the photonic reservoir (red circles). Statistics result from 5 independent test sets of $2^{18}$ bits.

We extend the investigation for the two data bit rate cases $R_1$ and $R_2$, by studying the dependence of the RC classification performance on the launched optical power in the transmission line on. Launched signals with low optical power, suffer by definition from low OSNR. For both bit rates, low optical power results in a poor RC post-processing performance (Fig. 7a and Fig. 7b). Only when the launched optical peak power is above 4dBm (7dBm), for the 56Gb/s (112Gb/s) data rate, the RC post-processing achieves the HD-FEC requirements in BER, for both $\theta$ conditions. At these power levels, Kerr nonlinearities and Brillouin scattering start to affect the transmitted signal. Nevertheless, by increasing the launched optical peak power up to 10 or 12dBm, we find a slight improvement in the decoding performance. RC shows a remarkable tolerance to these nonlinear effects, supporting a beneficially higher OSNR at the receiver end. This illustrates that the RC processing compensates not only for the chromatic dispersion, but for the introduced nonlinearities as well. The achieved BER is more than one order of magnitude lower, compared to our benchmark method (Fig. 7a and Fig. 7b, dashed lines).

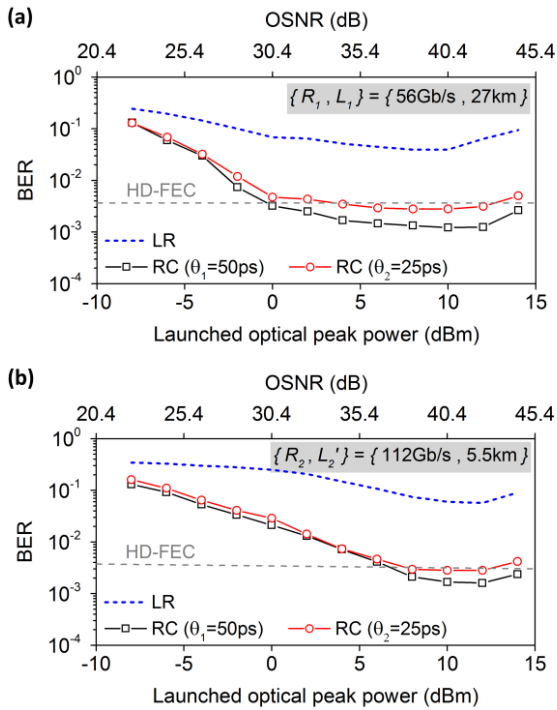

FIGURE 7. BER of the recovered PAM-4 data stream after transmission as a function of the launched optical peak power, for (a) {$R_1,L_1$} and (b) {$R_2,L_2$}. The launched optical peak power defines the signal's OSNR. LR: Linear regression on the transmission signal. RC: Linear regression on the reservoir output.

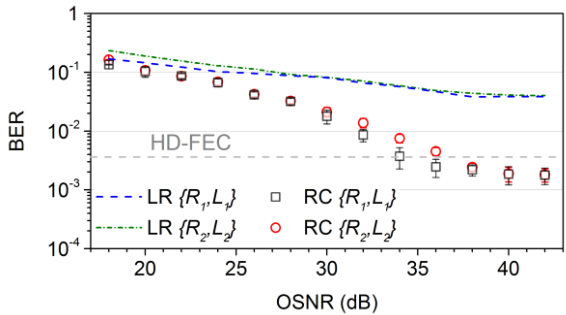

FIGURE 8. BER of the recovered PAM-4 data stream after transmission as a function of the OSNR for {$R_1,L_1$} and {$R_2,L_2$}. $\theta$=25ps. Statistics emerge from 5 independed test sets of $2^{18}$ bits. LR: Linear regression on the transmission signal. RC: Linear regression on the reservoir output.

Finally, we analyze the OSNR requirements for the given conditions and the transmission properties for the two bit rates. For a selected level of launched peak optical power – here 12dBm – we introduce different levels of optical noise from an optical amplifier into the channel to tune the OSNR of the received signal. The use of RC improves decoding for OSNR values above 20dB incrementally and saturates only above 40dB (Fig. 8). The obtained dependence is similar for both transmission systems. However, the {$R_1,L_1$} system exhibits a ~3dB gain in OSNR compared to the {$R_2,L_2$} system at the HD-FEC BER threshold.

## IV. EXPERIMENTAL PROOF OF CONCEPT

The experimental validation of the performance of our photonic reservoir uses the configuration in Fig. 2b. The signal $m(t) \cdot s_{norm}(t)$ to be processed by the photonic reservoir is applied to the MZM by using a 20GSa/s arbitrary waveform generator (AWG) with 10-bit resolution. In this investigation we do not consider pre-compensation of the MZM nonlinearity. The reservoir output $s_{rc}(t)$ is recorded with a 16GHz 80GSa/s real time oscilloscope.

The difference of the experimental implementation compared to the numerically simulated reservoir is the much longer delay of the feedback loop ($\tau$ = 66ns) due to the fiber-based setup. This length allows the definition of 1320 virtual nodes per delay, when considering $\theta$ = 50ps. However, in our experiment we connect the input signal to only N = 32 nodes – equal to the number of nodes considered in the simulated system – the responses of which have been used for the computation process. All remaining virtual nodes are not used. Moderate feedback conditions of the delayed reservoir are considered so that an optimal BER performance is obtained from the classification task. Specifically, this is achieved when the optical attenuation (ATT) in the feedback loop (Fig. 2b) is set at 18dB. Frequency detuning $\Delta f$ is set to 0GHz. These conditions are equivalent to the numerically estimated ones in Fig. 3b that lead to optimal performance. They also agree with the parameter space identified in the extensive dynamical study of [32] where optimal decoding efficiency is achieved.

Nevertheless, a direct comparison between the results obtained from numerical modelling and experimental measurements is not straightforward. In our experimental configuration, several sources of noise, non-ideal spectral response of individual hardware components and slight temperature variations affect the final evaluation of the system. Thus, we evaluate transmission links with slightly shorter distances in order to reach the HD-FEC limit for BER: $L_1'$=21km for the $R_1$ and $L_2'$=4.6km for the $R_2$ encoding rate. We consider a launched power in the transmission link of 10dBm. This power level leads to optimum BER level of the decoding process, according to our numerical results (Fig. 7). For the case of {$L_1',R_1$}, a linear classifier trained on the input signal from transmission $s(t)$ provides a BER level of $5.4 \cdot 10^{-3}$ (Fig. 9a, dashed line). The conversion of this numerical input signal into an actual electrical signal – via the AWG – has its own limitations that affect the final performance; the same linear classifier trained on the electrical input signal results in a BER performance which is now considerably worse (Fig. 9a, red dots). Moreover, there is a dependence of the BER on the detection signal-to-noise ratio (SNR), as shown in Fig. 9. The different SNR values originate from varying the number of averages we apply for the same repeatred detection signals. The averaging, which is performed in the oscilloscope, allows for an indirect increase of the measurement resolution. A single-shot measurement results in a SNR of 21dB, while performing 256 signal averages, the SNR increases up to 39dB. But even for the highest obtained SNR, the achieved BER of the AWG output signal ($1.8 \cdot 10^{-2}$) does not converge to the BER level of the initial numerical input. By feeding the AWG electrical output to the photonic reservoir, we obtain the reservoir responses $s_{rc}(t)$ and we train again the same linear classifier. The RC yields an improved performance compared to training on the directly detected output of the AWG (Fig. 9a, blue triangles), regardless of the SNR level. Moreover, for high SNR values (>35dB), the BER level after the RC processing is even lower than the one obtained for the numerically simulated input, and also below the HD-FEC BER limit. For the case of $\{L_2',R_2\}$ we find a similar performance and dependence on the SNR (Fig. 9b).

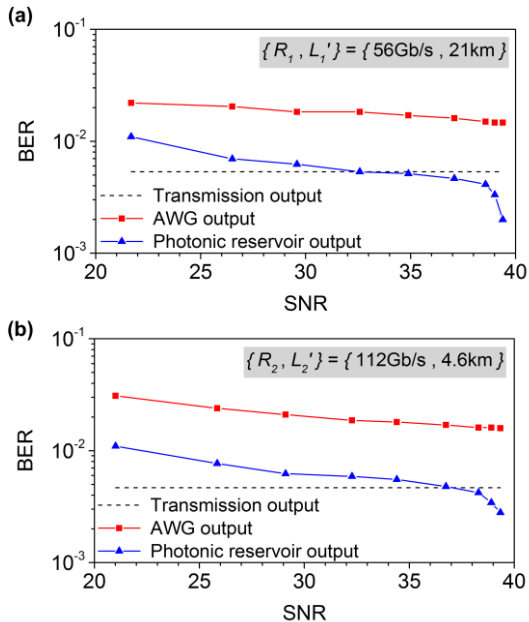

**FIGURE 9.** BER performance of the linear classifier on a PAM-4 data stream after transmission (dashed line), after electrical output conversion through an AWG (red dots) and after the photonic reservoir (blue triangles), versus the SNR of the detection system, for (a) $\{R_1,L_1\}$ and (b) $\{R_2,L_2\}$.

## V. CONCLUSIONS

Photonic RC is demonstrated to be an efficient and promising post-processing technique for transmission signals that have undergone complex nonlinear distortions. It allows the design of transmission systems that incorporate higher launched optical power in the fiber than usual and benefit from the higher OSNR. In the investigated scenarios of 56Gb/s and 112Gb/s encoding rates, the achieved communication distances are well above the ones described in the newly established IEEE protocol for short-reach fiber transmission, even though the latter refers to the wavelength region of 1.3µm where chromatic dispersion is minimized. The obtained performance can compete with the results using DSP in equivalent transmission systems [12]. Still, there are a number of future challenges for this type of hardware processing when considering ultrafast optical communication signals. Alternative or complementary approaches to the time-multiplexing photonic RC presented here need to be explored in order to establish real-time operation of this method. Furthermore, photonic realizations of data regression techniques will enable full photonic RC implementations for ultrafast signal processing.

## ACKNOWLEDGMENT
The authors are grateful to C. R. Mirasso, M. C. Soriano, S. Ortin for helpful discussions and P. Massuti-Ballester for technical support.

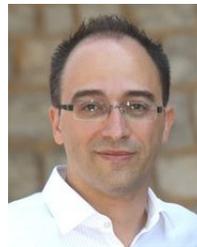

**APOSTOLOS ARGYRIS** received the B.Sc. degree in physics from the Aristotle University of Thessaloniki, Greece, in 1999, the M.Sc. degree in Microelectronics & Optoelectronics from the University of Crete, Greece in 2001 and the Ph.D. degree from the Department of Informatics and Telecommunications, National & Kapodistrian University of Athens (NKUA), Greece, in 2006, where he continued working as a research scientist until 2016.

He was an Adjunct Lecturer in the Departments of Computer Engineering, Telecommunications and Networks and Informatics of the University of Thessaly, Greece, from 2006 to 2009 and from 2014 to 2015. In 2016 he joined the Institute for Cross-Disciplinary Physics and Complex Systems – IFISC (CSIC-UIB) in Spain as a Marie-Skłodowska Curie Fellow. He has participated in 12 research projects, funded by the European Union and the national Greek / Spanish research authorities. He has published more than 80 papers in peer-reviewed journals and conference proceedings, while he is the main author of 3 book chapters. His research interests include nonlinear dynamics in semiconductor lasers, photonic devices and coupled systems, optical communication systems, chaotic encryption, optical sensing and machine learning techniques.

Dr. Argyris was distinguished as one of the world's top young innovators from the Technology Review magazine of Massachusetts Institute of Technology, Boston, U.S.A. by receiving the "TR35 Award" in 2006. The same year he was awarded the "Ericsson Telecommunications Award of Excellence" for his PhD thesis.

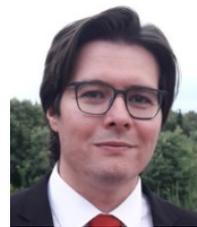

**JULIAN BUENO** received the B.Sc. degree in physics from the Universitat de les Illes Balears (UIB) in 2013 and the M.Sc. degree in Complex Systems from the Institute for Cross-Disciplinary Physics and Complex Systems (IFISC) in 2014.

He is currently working towards his PhD defense, while he recently joined the Institute of Photonics, of the University of Strathclyde, U.K.. Before graduating he also studied for one year at the Vrije Universiteit Brussel (VUB). His research interests focuses on nonlinear photonics, complex dynamics and implementations of neural networks in photonic hardware.

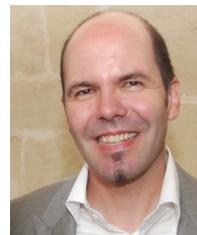

**INGO FISCHER** received his Diploma and Ph.D. degrees in physics from Philipps University Marburg (Germany) in 1992 and 1995, respectively.

He was at the Technical University of Darmstadt (Germany) from 1995 to 2004, and at the Vrije Universiteit Brussel (Belgium) from 2005 to 2007. In 2007, he became a Full Professor in photonics at Heriot-Watt University, Edinburgh (U.K.). Since 2009, he has been a Professor at the Institute for Cross-Disciplinary Physics and Complex Systems, a joint center of the Spanish National Research Council (CSIC) and the University of the Balearic Islands (UIB), Palma de Mallorca (Spain). His current research interests include nonlinear photonics and neuro-inspired information processing, and in particular, the emission properties and dynamics of modern photonic sources, photonic computing, coupled laser systems, synchronization of lasers, and utilization of chaos. He has authored or co-authored more than 100 papers in peer-reviewed journals and has been PI of 19 projects financed by European Union, German Research Foundation, Spanish Ministry and other institutions.

Dr. Fischer received the Research Prize of the Adolf-Messer Foundation in 2000, and the first Hessian Cooperation Prize of the Technology Transfer Network in 2004.